\documentclass[epj]{webofc}
\usepackage[utf8]{inputenc}
\usepackage[varg]{txfonts}   
\usepackage{booktabs}
\usepackage{xcolor}
\usepackage{url}
\definecolor{darkred}{rgb}{0.4,0.0,0.0}
\definecolor{darkgreen}{rgb}{0.0,0.4,0.0}
\definecolor{darkblue}{rgb}{0.0,0.0,0.4}
\usepackage[bookmarks,linktocpage,colorlinks,
    linkcolor = darkred,
    urlcolor  = darkblue,
    citecolor = darkgreen]{hyperref}
\usepackage{subfigure}
\usepackage{tikz}

\usepackage{listings}

\usepackage{multicol}
\usepackage{verbatim}

\definecolor{commentgreen}{RGB}{2,112,10}

\lstdefinestyle{mystyle}{
    escapechar={!},
    language=C++,
    keywordstyle=\color{purple},
    commentstyle=\color{commentgreen},
    stringstyle=\color{magenta},
    numberstyle=\tiny\color{gray},
    basicstyle={\scriptsize\ttfamily},
    frame=single,
    breaklines=true,
    captionpos=b,
    keepspaces=true,
    numbersep=5pt,
    showspaces=false,
    showstringspaces=false,
    showtabs=false,
    tabsize=2,
    firstnumber=1,
    numberfirstline=true,
    otherkeywords={OFFLOAD,\_\_global\_\_,pragma},
    morekeywords=[2]{OFFLOAD,\_\_global\_\_},
    morekeywords=[3]{loop, independent, target, teams, device, pragma, acc, omp, parallel, copyin, copyout, mapto, mapfrom, map, distribute, for},
}

\lstdefinestyle{mystyle2}{
    escapechar={!},
    language=C++,
    keywordstyle=\color{purple},
    commentstyle=\color{commentgreen},
    stringstyle=\color{magenta},
    numberstyle=\tiny\color{gray},
    basicstyle={\scriptsize\ttfamily},
    breaklines=true,
    captionpos=b,
    keepspaces=true,
    numbersep=5pt,
    showspaces=false,
    showstringspaces=false,
    showtabs=false,
    tabsize=2,
    firstnumber=1,
    numberfirstline=true,
    otherkeywords={OFFLOAD,\_\_global\_\_,pragma},
    morekeywords=[2]{OFFLOAD,\_\_global\_\_},
    morekeywords=[3]{loop, independent, target, teams, device, pragma, acc, omp, parallel, copyin, copyout, mapto, mapfrom, map, distribute, for},
}

\wocname{EPJ Web of Conferences}
\woctitle{Lattice2017}
\begin{document}
%
\selectlanguage{english}
\title{
Performance Portability Strategies for Grid C++ Expression Templates}
\author{%
\firstname{Peter A.} \lastname{Boyle}\inst{1}  \and
\firstname{M.A.} \lastname{Clark}\inst{2} \and
\firstname{Carleton}  \lastname{DeTar}\inst{3} \and 
\firstname{Meifeng} \lastname{Lin}\inst{4}\fnsep\thanks{Speaker, email: mlin@bnl.gov}  \and 
\firstname{Verinder} \lastname{Rana}\inst{4} \and 
\firstname{Alejandro} \lastname{Vaquero Avil\'es-Casco}\inst{3}
}
\institute{%
Higgs Centre for Theoretical Physics,  School of Physics \& Astronomy, University of Edinburgh, EH9 3FD, UK
\and
NVIDIA Corporation, Santa Clara, CA 95050, USA
\and 
Department of Physics and Astronomy, University of Utah, Salt Lake City, UT 84112, USA
\and
Computational Science Initiative, Brookhaven National Laboratory, Upton, New York 11973, USA
}
\abstract{%
One of the key requirements for the Lattice QCD Application Development as part of the US Exascale Computing Project is performance portability across multiple architectures. Using the Grid C++ expression template as a starting point, we report on the progress made with regards to the Grid GPU offloading strategies. We present both the successes and issues encountered in using CUDA, OpenACC and Just-In-Time compilation. Experimentation and performance on GPUs with a SU(3)$\times$SU(3) streaming test will be reported. We will also report on the challenges of using current OpenMP 4.x for GPU offloading in the same code.
}
\maketitle
\section{Introduction}\label{sec:intro}

The USQCD Collaboration and its partners\footnote{As of Lattice 2017, our partners included University of Edinburgh, University of Illinois, NVIDIA and Stony Brook University.} are engaged in the application development (AD) efforts~\cite{thiscontrib396} as part of the US Exascale Computing Project~\cite{ecp} to redesign a software stack for lattice QCD simulations for the anticipated exascale computing systems becoming available in a few years. While we do not know the exact configurations of the exascale systems, the architectures are expected to be diverse, and consist of heterogeneity, complex memory hierarchies and multiple levels of parallelism.  One of the key requirements for the USQCD exascale data-parallel applications programming interface (API) is \textit{performance portability}. That is, the ability to run the same code on different architectures without losing too much performance. The Grid~\cite{Boyle:2016lbp} C++ data-parallel library is being considered as one of the candidate frameworks on top of which the data-parallel API will be developed. Therefore an investigation of the ease of portability in Grid has been performed.

Grid is a data-parallel library for lattice QCD. It uses C++11 features and has its own expression template engine. Grid has been optimized for CPU-based systems, including Intel Xeon CPUs, Intel Knights Corner and Knights Landing processors, and the IBM BG/Q systems. Its use of expression template engine and complex data structures makes it challenging to port to GPUs. The initial effort to port Grid to GPUs started in summer 2016 when a subset of the authors implemented a GPU version of the stripped-down Grid expression template engine using a simple CUDA kernel~\cite{Boyle:2017wul}.  Since then, the QCD ECP AD team has been exploring more extensively different portability approaches for Grid. Since the whole Grid library is quite complicated and has tens of thousands of lines of code, we again started out with a simple prototype code that contains the Grid expression template engine for proof of concept, and develop examples to include more realistic use cases in QCD simulations. This paper summarizes the findings and roadblocks in the various approaches.

Our main considerations are \emph{ease of use} and \emph{performance}:

\begin{itemize}

\item{Ease of use.} The portability requirement of the USQCD exascale software API means we do not want to rewrite the whole library in order to port it to GPUs. Instead, we want to reuse as much code as we can, and the ideal portability technique
would require only a few trivial modifications. 

\item{Performance.} We aim to develop software optimized for the next generation of supercomputers, capable of computing at the exascale level. The portable code should perform well in all next-generation platforms, in order to stay competitive.
\end{itemize}

To meet both of these two requirements can be complicated, and compromises may be needed. Code with high portability can be accepted if the performance loss is minor. A small rewrite of a few key parts of the library might be acceptable. Hence we must look for a method that will achieve the most balanced result. To this end, we have studied OpenACC, OpenMP, Just-In-Time compilation and CUDA as part of our performance portability research, which will be described in Section~\ref{sec:pm}. Following a proof-of-concept test in Section~\ref{sec:et}, we show the performance of a lattice-wide SU(3)$\times$SU(3) test in Sections~\ref{sec:su3} and \ref{sec:vsu3}, first with a scalar data layout and then with a vector data layout. Section~\ref{sec:summary} contains the summary and outlook of the project.

\section{Technical Approaches} \label{sec:pm}

\subsection{OpenACC}

OpenACC is a standard of compiler directives that allows offloading computationally intensive tasks to an accelerator. In C/C++ code, OpenACC statements are introduced as
\begin{lstlisting}[style=mystyle2]
#pragma acc ...
\end{lstlisting}
where the \ldots should be replaced with one of the multiple OpenACC directives available. These statements indicate parallelizable regions or data movement from/to the accelerator, and the compiler generates the necessary compute and data movement kernels based on the directives. In this respect,
OpenACC is quite similar to  the familiar OpenMP shared-memory programming model. However, to offload a compute region to the GPU using OpenACC, we also need to be concerned about data movement, whereas in the OpenMP threading model, the programmer is mostly concerned about how to distribute the threads.  

OpenACC loops can be divided into three levels: gangs, workers and vectors. A gang represents a group of workers, and a worker a group of vectors. They exist to accommodate different degrees of parallelism existing in a GPU (there is a grid of blocks, and each block contains several threads). They can also be mapped to multicore vector architectures. Users can also choose to use descriptive directives such as \texttt{kernels} which will give the compiler the freedom to analyze the code and generate compute kernels as it sees fit.  For a more detailed introduction to OpenACC, refer to ~\cite{Farber20171}.

OpenACC, as  a directive-based approach, is easy to introduce in any existing code, and satisfies our ``ease of use'' requirement. It also allows compilation for a large number of targets, such as NVIDIA GPUs, AMD GPUs and multicore architectures, making it a highly portable approach. However, the generated kernels are dependent on the compiler, giving users limited control over performance. 
In addition, while C code is easy to port to GPUs using OpenACC, porting C++ code is often non-trivial due to complicated data structures that would make it hard to manage the data movement between CPU and GPU without the support of deep copy~\cite{Farber20171}. PGI compiler's support for the unified virtual memory (UVM) greatly alleviates the difficulty with the data management for C++ code, but other issues such as the lack of support for C++ standard template library (STL) persist.

\subsection{OpenMP 4.5}
Recently in OpenMP 4.5, the concept of fork-join parallelism has been extended to accelerator devices. Computationally intensive code can now be offloaded onto the accelerators thanks to this feature. If multiple accelerators are connected to the host, OpenMP allows one to take advantage of them by specifying their device IDs. To execute on the device, OpenMP uses the \texttt{\#pragma omp target} construct, which is similar to the OpenACC \texttt{\#pragma acc parallel} construct. The \texttt{\#pragma omp teams} construct starts a team of threads and begins parallel execution. The data directives in OpenMP are given by the \texttt{map} clause which further gets decorated using \texttt{to}, \texttt{from}, \texttt{tofrom} which essentially are similar to \texttt{copyin}, \texttt{copyout}, \texttt{copy} in OpenACC. In addition, the concepts of \texttt{gang}, \texttt{worker}, \texttt{vector} in OpenACC correspond to  \texttt{teams}, \texttt{distribute parallel for}, \texttt{distribute simd} in OpenMP.

OpenMP is similar to OpenACC in the sense that they are both relying on the compiler to generate the compute kernels. However, OpenMP faces more challenges than OpenACC for C++ code due to the lack of UVM support in the current compiler implementations~\cite{Karlin2016,omp-hack}. 

\subsection{CUDA} \label{sec:cuda}
CUDA is NVIDIA's API for GPU programming. Since it is vendor-specific, it does not provide true portability. Given that NVIDIA's GPUs are the main-stream GPU architecture for high performance computing, we consider this an acceptable approach. It is also necessary to write some CUDA kernels, which will cause some code branching and potentially enlarge the code size significantly. In addition, all functions called by the CUDA kernels need to be decorated with \texttt{\_\_device\_\_} qualifier and with \texttt{\_\_host\_\_} as well if it is also compiled for CPU. That may become too tedious for a large code base\footnote{Future versions of CUDA may allow for \texttt{\_\_host\_\_ \_\_device\_\_} namespaces, which can greatly simplify this approach.}. However, it is a mature GPU programming model that offers good performance on NVIDIA GPUs, and its C++ support has been steadily improving.  This approach is also interesting because Grid has important areas of code such as the stencil operators that would also need to be offloaded to the GPUs and may
be simple to do with CUDA. 

\subsection{Just-In-Time (JIT) Compilation and Jitify}
An alternative to using CUDA and compiler directives is to use the CUDA runtime compiler \texttt{nvrtc} to perform ``Just-In-Time" (JIT) compilation on demand. Since \texttt{nvrtc} is a pure GPU compiler, it is not as constrained as the off-line heterogeneous compiler, and can be forced to interpret all encountered functions as \texttt{\_\_device\_\_} functions, eliminating the need to decorate the functions \texttt{\_\_host\_\_ \_\_device\_\_} necessary for explicit CUDA programming. 


Jitify~\cite{jitify-git} is a new C++ header library developed at NVIDIA that provides a simple front-end to \texttt{nvrtc}. It uses C++11 features to present simple single lambda-style launch syntax, and can choose GPU or CPU as runtime execution policy, where CPU code is compiled off-line and GPU code is compiled during runtime. Jitify removes all CUDA-specific extensions from user code, but the user can still access them if needed. However, its just-in-time compilation nature makes it hard to detect problems before program execution. In addition, all kernel functions need to be known in the header files.

\section{Porting the Grid Expression Template Engine}\label{sec:et}
In this section we demonstrate the proofs of concept of using the above approaches in a simple stripped-down example code. Listing~\ref{lst:ET-cpu}  shows the key components of the CPU version of the stripped-down expression template in Grid, where the main computation occurs in the highlighted \texttt{for} loop. 

\begin{lstlisting}[style=mystyle, caption=Grid expression template, label=lst:ET-cpu]
 template <typename Op, typename T1,typename T2> inline Lattice<obj> & operator=(const  LatticeBinaryExpression<Op,T1,T2> expr)
    {
      int _osites=this->Osites();
       !\colorbox{yellow}{for(int ss=0;ss<\_osites;ss++)\{}!
        _odata[ss] = eval(ss,expr);
      }
      return *this;
    }
  };
\end{lstlisting}

The expression template engine will allow for evaluations of arbitrary expressions such as the ones shown below, where \texttt{v1, v2} and \texttt{v3} are lattice-wide arrays of the \texttt{double} type: 
\begin{lstlisting}[style=mystyle,caption=Example composite expressions., label=lst:et-test]
  Lattice<double> v1(&grid);
  Lattice<double> v2(&grid);
  Lattice<double> v3(&grid);
  v1=1.0;
  v2=2.0;
  v3=v1+v2;
  v3=v1+v2+v1*v2;
\end{lstlisting}

For each of the approaches, the code changes needed to get the expressions to be evaluated on the GPU are shown in Table~\ref{tab:code-changes}. All of our implementations assume UVM support, either through the \texttt{cudaMallocManaged} allocator for CUDA and Jitify, or through \texttt{-ta=managed} compile flag for OpenACC with the PGI compiler. The OpenMP implementation did not work due to its lack of deep-copy or UVM support. 

\begin{table}[htbp]
\centering
\begin{tabular}{|l|l|}
\hline
OpenACC & 
{\begin{lstlisting}[style=mystyle2]
#pragma acc parallel loop independent copyin(expr[0:1]) 
 for(int ss=0;ss<_osites;ss++){
    _odata[ss] = eval(ss,expr);
 }
\end{lstlisting} }

\\ \hline
OpenMP & 
{\begin{lstlisting}[style=mystyle2]
#pragma omp target device(0) map(to: expr) map(tofrom:_odata[0:_osites])
  {
  #pragma omp teams distribute parallel for
     {
     for (int i=0; i<_osites; i++)
        _odata[ss] = eval(ss,expr);
     }
   }
\end{lstlisting} }
\\ \hline
Jitify & 
{\begin{lstlisting}[style=mystyle2]
parallel_for(policy, 0, _osites, 
              JITIFY_LAMBDA( (_odata,expr), 
              _odata[i]=eval(i,expr); ));
\end{lstlisting}}
\\ \hline
CUDA & 
{\begin{lstlisting}[style=mystyle2]
 template<class Expr, class obj> __global__
  void ETapply(int N,obj *_odata,Expr Op)
  {
    int ss = blockIdx.x;
    _odata[ss]=eval(ss,Op);
  }
 LatticeBinaryExpression<Op,T1,T2> temp = expr;
 ETapply< decltype(temp), obj > <<<_osites,1>>>((int)_osites,this->_odata,temp);
\end{lstlisting}}
\\ \hline
\end{tabular}
\caption{Comparison of compute kernels in different GPU programming models. }\label{tab:code-changes}
\vspace{-2mm}
\end{table}





\section{SU(3)$\times$SU(3) With Scalar Data Layout}\label{sec:su3}
In this section we demonstrate implementations using CUDA, OpenACC and Jitify of an SU(3)xSU(3) streaming test. The starting point is a templated $N\times N$ matrix class which is then used as the data type for the expression template, i.e. replacing \texttt{double} in Listing~\ref{lst:et-test} with \texttt{Su3f} (for single-precision complex $3\times 3$ matrices).  
\begin{lstlisting}[style=mystyle, caption=SU(3)$\times$SU(3) test code snippet.]
  Lattice<Su3f> z(&grid);
  Lattice<Su3f> x(&grid);
  Lattice<Su3f> y(&grid);
  for(int i=0;i<Nloop;i++) {
        z=x*y;
  }
\end{lstlisting}

In this n\"aive implementation, the  lattice-wide arrays of complex $3\times 3$ matrices are not vectorized, unlike in the actual Grid implementation where vectorized layout is built-in (see Section~\ref{sec:vsu3}). This Array of Structure (AoS) data order limits its achievable performance on the GPU due to the lack of memory coalescence. To boost performance we used the \texttt{coalesced\_ptr} class~\cite{cptr} that  automatically break up AoS into AoSoAoS order to ensure coalesced thread access.  

\begin{figure}[htbp]
\begin{minipage}{0.49\textwidth}
\centering
\includegraphics[width=\textwidth]{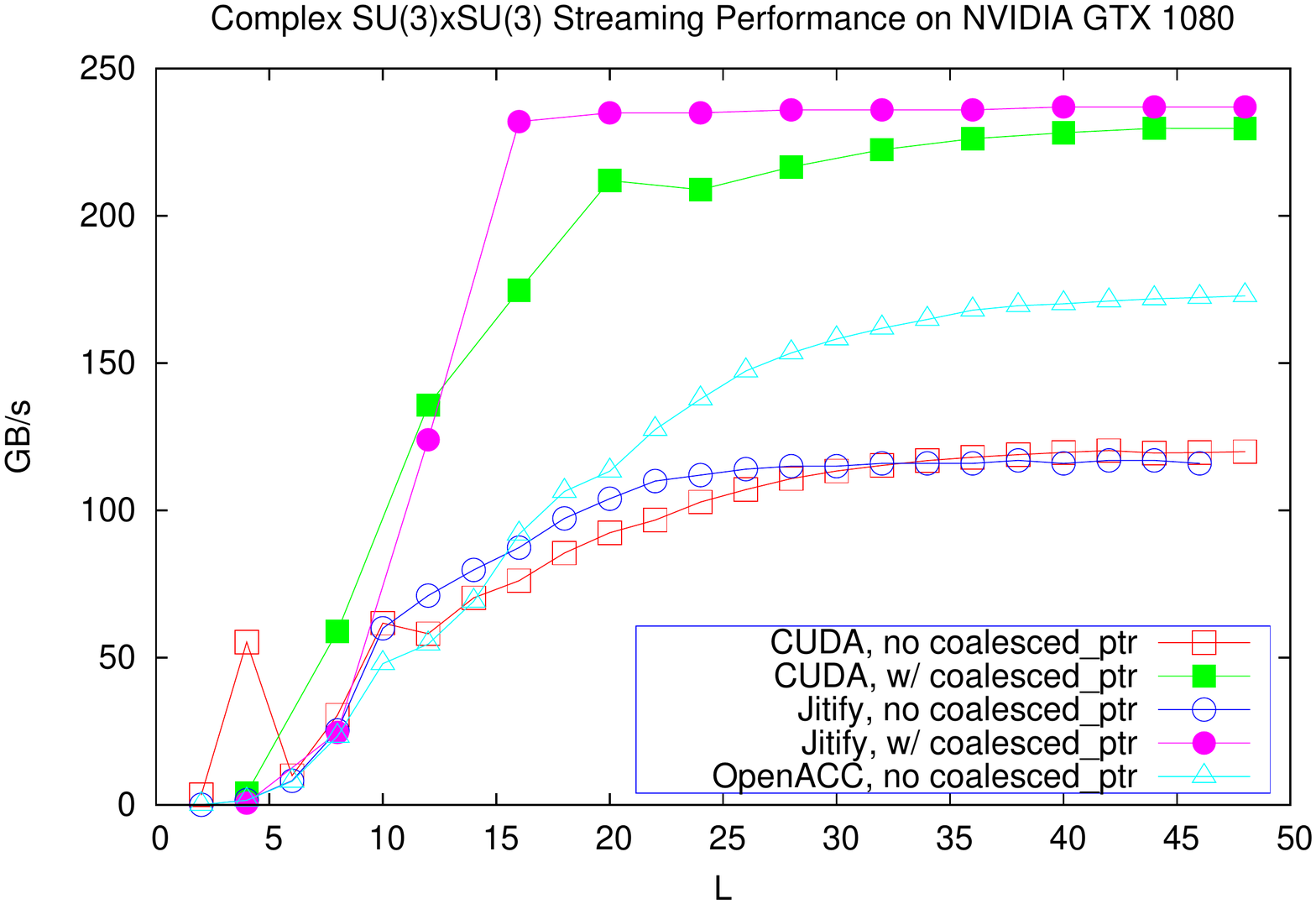}
\vspace{-1cm}
\caption{Complex SU(3)$\times$SU(3) streaming bandwidth on NVIDIA GTX 1080 for implementations with OpenACC, CUDA and Jitify.} \label{fig:su3_GBs} 
\end{minipage}
\hfill
\begin{minipage}{0.49\textwidth}
\centering
\includegraphics[width=\textwidth]{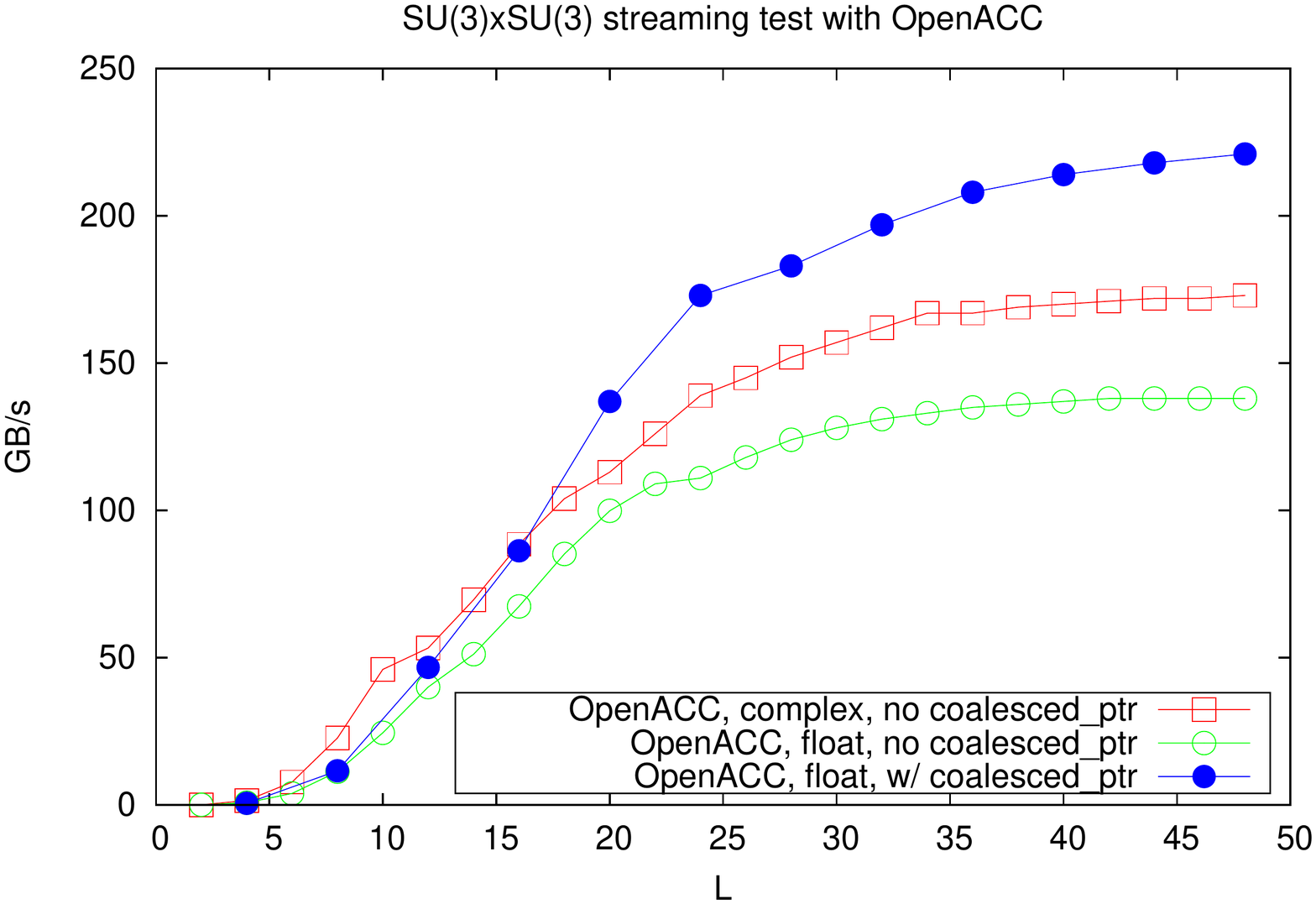}
\vspace{-1cm}
\caption{Comparison of OpenACC SU(3)$\times$SU(3) streaming bandwidth with and without coalesced pointer.} \label{fig:su3_openacc}
\end{minipage}
\vspace{-2mm}
\end{figure}

We performed our tests on a workstation with NVIDIA GTX 1080 and quad-core Intel i7 CPU. The maximum bandwidth on the GTX 1080 is 288 GB/s, with maximum sustainable bandwidth of around 240 GB/s. Shown in Figure~\ref{fig:su3_GBs} are the bandwidth results with our CUDA, Jitify and OpenACC implementations. Without the use of the \texttt{coalesced\_ptr} class, CUDA and Jitify could obtain only about half of the maximum sustainable bandwidth ($\sim$ 120 GB/s), while OpenACC outperforms both of them. When \texttt{coalesced\_ptr} is used, both CUDA and Jitify saturated the memory bandwidth, with CUDA under-performing Jitify at large memory sizes and outperforming at small memory sizes, likely due to the different thread and block number choices in these two approaches. OpenACC with  \texttt{coalesced\_ptr} did not work with complex data types due to PGI compiler bugs, but with real data type (\texttt{float}), OpenACC with \texttt{coalesced\_ptr} was also able to deliver reasonable memory bandwidth, as shown in Figure~\ref{fig:su3_openacc}. 

\section{SU(3)$\times$SU(3) With Vector Data Layout} \label{sec:vsu3}
While using the \texttt{coalesced\_ptr} class allowed us to achieve very good bandwidth in our SU(3)$\times$SU(3) tests, it is interesting to see if we can obtain coalesced access with Grid's native vector data layout that is optimized for the SIMD processors.  To achieve efficient vectorization Grid splits the \texttt{Lattice} object on each compute node into $N_{sub}$ virtual nodes, where $N_{sub}$ is equal to the SIMD length of the processors divided by the floating point precision. The splitting of the \texttt{Lattice} object in SIMD lanes occurs at the level of the basic data types such as \texttt{complex} or \texttt{float}. Any higher-level data types built on these fundamental data types will be vectorized as well.


On the CPU, parallelization on SIMD lanes is straightforward because the bulk points in different SIMD lanes are independent from each other and can be computed at the same time.  On the GPU, 
this Array-of-Structures-of-Vectors/Arrays  (AoSoV, or AoSoA) data layout allows for an efficient mapping to GPU threads. For example, each thread block can process one or several vectors, and the threads can operate on different SIMD lanes of each vector. This approach should achieve coalesced access if the threads in a warp are forced to read data in consecutive memory. 

The main obstacle in this approach is the fact that our \texttt{Lattice} object hosts vectorized data types. In the particular case of our SU(3)$\times$SU(3) test, each element of the matrix is actually a vector of real or complex numbers corresponding to different virtual nodes. If we apply the expression template as before without any changes, each GPU thread will operate on the vector objects, rather than one element of the SIMD vector, which is the complete opposite of what we would want. Instead, we need to extract the scalar data from the vector data types before the GPU computation, and then merge them back to form the vector objects after, as shown below:
  $$\begin{pmatrix}
   {\color{red} a^{11}_1}{\color{green} a^{11}_2}{\color{blue} a^{11}_3}{\color{orange} a^{11}_4} & {\color{red} a^{12}_1}{\color{green} a^{12}_2}{\color{blue} a^{12}_3}{\color{orange} a^{12}_4} & {\color{red} a^{13}_1}{\color{green} a^{13}_2}{\color{blue} a^{13}_3}{\color{orange} a^{13}_4} \\
         & & & \\
   {\color{red} a^{21}_1}{\color{green} a^{21}_2}{\color{blue} a^{21}_3}{\color{orange} a^{21}_4} & {\color{red} a^{22}_1}{\color{green} a^{22}_2}{\color{blue} a^{22}_3}{\color{orange} a^{22}_4} & {\color{red} a^{23}_1}{\color{green} a^{23}_2}{\color{blue} a^{23}_3}{\color{orange} a^{23}_4} \\
         & & & \\
   {\color{red} a^{31}_1}{\color{green} a^{31}_2}{\color{blue} a^{31}_3}{\color{orange} a^{31}_4} & {\color{red} a^{32}_1}{\color{green} a^{32}_2}{\color{blue} a^{32}_3}{\color{orange} a^{22}_4} & {\color{red} a^{33}_1}{\color{green} a^{33}_2}{\color{blue} a^{33}_3}{\color{orange} a^{33}_4} 
    \end{pmatrix}\quad\longleftrightarrow\quad
    \begin{matrix}
      \begin{pmatrix} {\color{red}  a^{11}_1} & \ldots \\ \vdots & \ddots\end{pmatrix} & \begin{pmatrix} {\color{green}  a^{11}_2} & \ldots \\ \vdots & \ddots\end{pmatrix} \\
      \textrm{Thread 1} & \textrm{Thread 2} \\
       & \\
      \begin{pmatrix} {\color{blue} a^{11}_3} & \ldots \\ \vdots & \ddots\end{pmatrix} & \begin{pmatrix} {\color{orange} a^{11}_4} & \ldots \\ \vdots & \ddots\end{pmatrix} \\
      \textrm{Thread 3} & \textrm{Thread 4} 
    \end{matrix}$$

Grid provides functions to perform the extraction/insertion of scalarized objects from/into vectorized ones. In terms of the expression template engine, we implemented the extract and merge functions so that each GPU thread processes one SIMD lane and each block gets the whole vector. 

The relevant code snippets are shown in Listing~\ref{lst:extract}.

\begin{lstlisting}[style=mystyle, caption=Extract and merge the vector objects in Grid., label=lst:extract]
 //Make each thread {eval} one element of the vector, extracted through {extractS}. 
    auto sD = extractS<obj,sObj>(arg._odata[ss], tIdx); 
    ...
 //Put the results back to form the vector again after {evalS}.
    auto sD = evalS(blockIdx.x,Op,threadIdx.x);
    mergeS(_odata[blockIdx.x], sD, threadIdx.x);
\end{lstlisting}
      
\begin{figure}[t]
  \begin{minipage}{0.49\textwidth}
  \centering
    \includegraphics[width=\textwidth,angle=0]{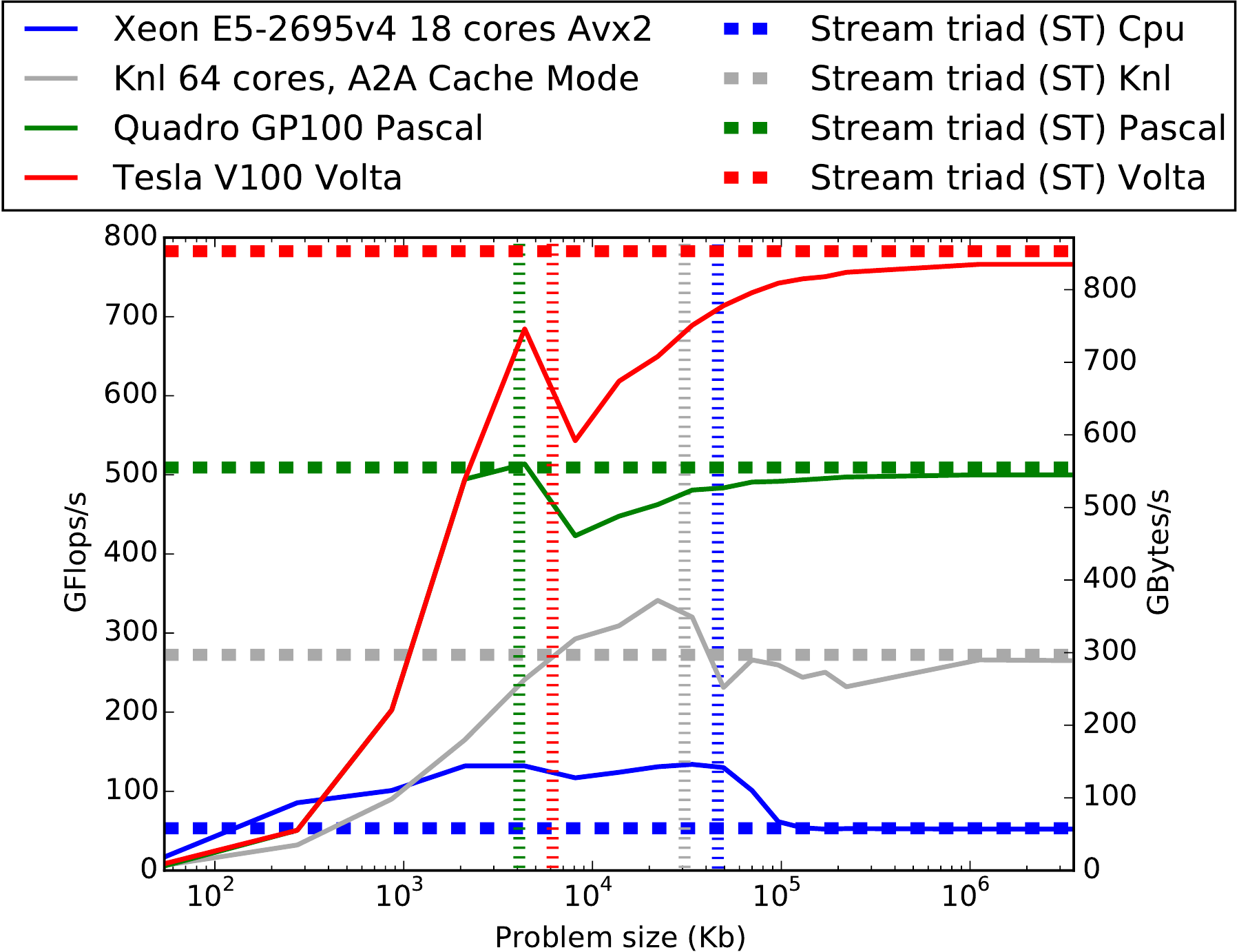}
    \caption{Performance of the miniapp in several architectures. The vertical lines indicate the cache size, the horizontal ones show the memory bandwidth obtained in a stream triad test. In all cases we can easily saturate the memory bandwidth of the machine.}\label{fig:vsu3-perf}
  \end{minipage}
  \hfill
  \begin{minipage}{0.49\textwidth}
    \centering
    \vspace{12mm}
    \includegraphics[width=\textwidth,angle=0]{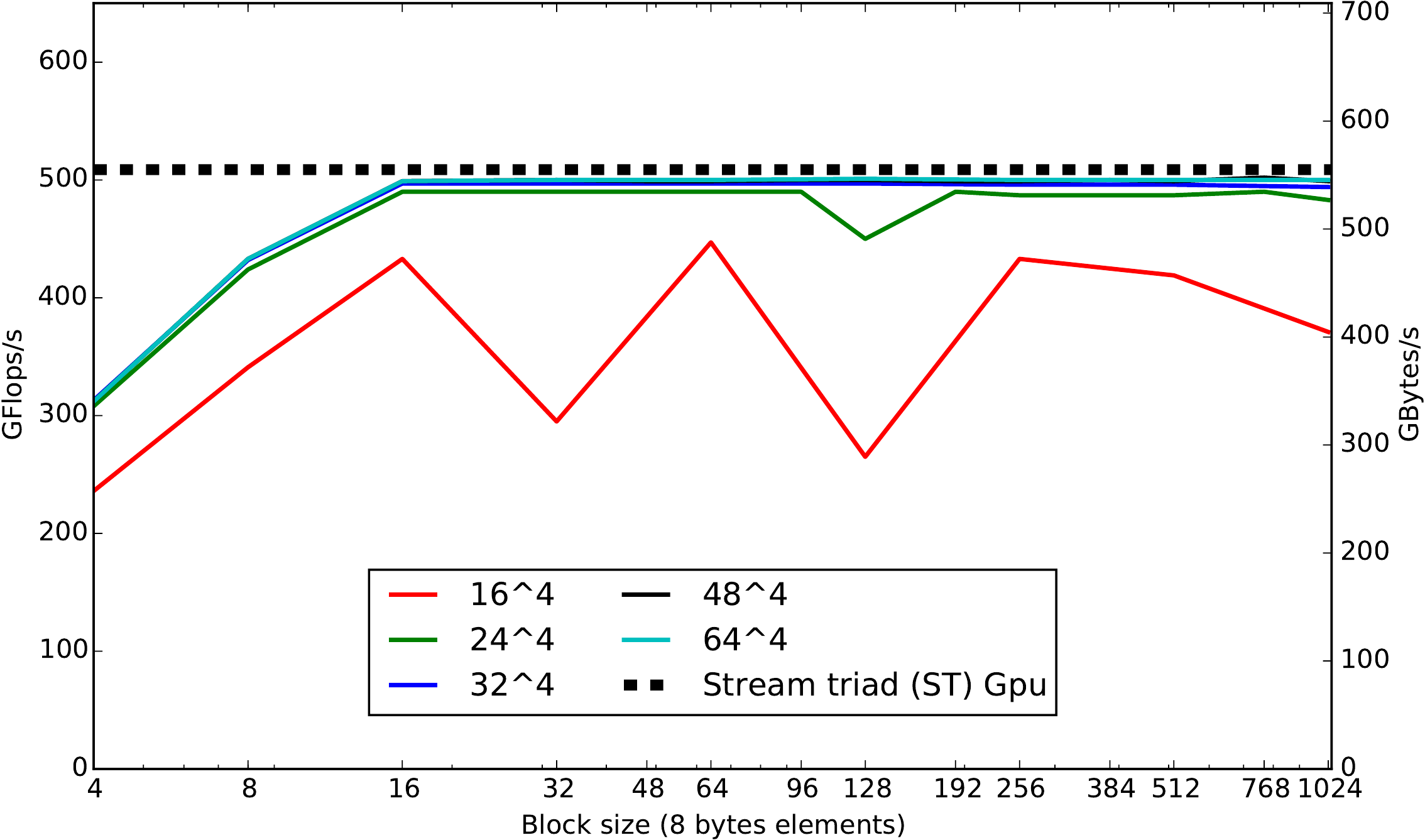}
    \caption{Performance of the miniapp as a function of the vector length in a Quadro GP100. The vector length is measured in terms of single precision complex numbers. For a vector of 16 complex numbers (twice the size of a KNL vector), performance  hits the maximum.}\label{fig:vsu3-blocklen}
  \end{minipage}
  \vspace{-5mm}
\end{figure}

With the same code we could run the SU(3)$\times$SU(3) tests on several different platforms, including the Intel Xeon E5-2695v4 ``Broadwell" processor at 2.1 GHz, the Intel ``Knights Landing" Xeon Phi(TM) CPU 7230 \@ 1.30GHz~\footnote{Booted in all-to-all cache mode which renders lower bandwidth than the quad cache mode.}, the NVIDIA Quadro GP 100 and Tesla V100 GPUs, the results of which are shown in Figure~\ref{fig:vsu3-perf}. The left axis shows the measured floating point performance, while the right axis shows the measured bandwidth. The horizontal dashed lines indicate the plateaued results from STREAM Triad tests on the same platforms, and the vertical lines indicate the cache sizes of these architectures. The cache effects are evident, and the fact that our results follow closely the STREAM triad results shows that our implementation indeed achieves the maximum memory bandwidth possible. 

The GPU performance may depend on the vector length of the data types, or in GPU language, the block size in the kernel call, as shown in Figure~\ref{fig:vsu3-blocklen}. The performance plateaus with a block size of 16 threads, or equivalently a vector length of 16 single-precision complex numbers. The large vector length required to achieve the best performance is not desirable as it puts more constraints on the lattice dimensions and vectorization choices. A solution is to create a two-dimensional block and assign several vectors per block. The $x$ coordinate of the thread index would run through the SIMD lanes, as before, but now the $y$ coordinate would run over the different vectors inside the block. This has not been implemented yet, but will be in the near future. 

\section{Summary and Outlook}\label{sec:summary}
We have investigated three approaches for performance portability in the context of Grid expression templates: OpenACC and/or OpenMP directives, CUDA and Jitify. First we demonstrated how we could offload the Grid expression template engine to GPUs in a simple test case. Then we did some performance studies in the SU(3)$\times$SU(3) streaming test, first with a naive data layout, then with Grid's vector data layout.  

OpenACC, in combination with PGI compiler's UVM support, can be used to offload Grid's expression template engine, and also delivers reasonable performance in our SU(3)xSU(3) streaming test. However, there are a number of compiler issues that potentially limit our ability to use OpenACC in a large code base. 

Jitify eliminates the need to decorate the device functions, has a simple user interface and can deliver good performance if used with coalesced pointer class. However, it is a relatively new JIT library and still under active development. Its usability in a large code still needs to be tested. 

Using CUDA to offload requires all the host functions also called in a GPU kernel to be decorated as \texttt{\_\_host\_\_ \_\_device\_\_}, which may become tedious and error-prone in a large code base. Another issue we encounter repeatedly is that using \texttt{std::complex} would cause the compiler to complain, likely due to the fact that the operators in std::complex are not decorated as \texttt{\_\_host\_\_ \_\_device\_\_} functions. This is a universal issue with the GPU offloading of the C++ STL. But in the example we have tested, CUDA shows as a viable option both in terms of ease of programming and performance, and is our programming model of choice for porting Grid to the GPU given the current state of the art. 

We have also demonstrated that Grid's SIMD-friendly vector data layout can be easily mapped to the GPU execution model and achieve good performance. To eliminate the need for a large vector length in order to achieve the best GPU performance, a two-dimensional thread block mapping may be necessary, which will be implemented in the future. We are in the process of porting the Dslash operator in Grid to GPUs, in which some routines that are outside of the Grid expression template engine, such as the Cshift operations, may require additional attention. 

\section*{Acknowledgments}
 The authors acknowledge many helpful discussions with the rest of the USQCD ECP team. This research was supported by the Exascale Computing Project (17-SC-20-SC), a collaborative effort of the U.S. Department of Energy Office of Science and the National Nuclear Security Administration. Part of this research was carried out at the Brookhaven Hackathon 2017. We thank Mathias Wagner (NVIDIA) for serving as our mentor at the hackathon. Brookhaven Hackathon is a collaboration between and used resources of Brookhaven National Laboratory, University of Delaware, Stony Brook University, and the Oak Ridge Leadership Computing Facility at the Oak Ridge National Laboratory. We gratefully acknowledge the use of the computing resources at Brookhaven National Laboratory and NVIDIA for our performance tests. 
\bibliography{lattice2017}

\end{document}